\documentclass[reprint]{revtex4-1}

\draft 

\usepackage{graphicx}
\usepackage{dcolumn}
\usepackage{bm}

\begin{document}

\title{Optical Yagi-Uda nanoantennas} 

\author{Ivan S. Maksymov, Isabelle Staude, Andrey E. Miroshnichenko, and Yuri S. Kivshar}

\email[Ivan S. Maksymov ]{mis124@physics.anu.edu.au}

\affiliation{Nonlinear Physics Centre and Centre for Ultrahigh Bandwidth Devices for Optical Systems (CUDOS), Research School of Physics and Engineering, The Australian National University, Canberra, ACT 0200, Australia}

\date{\today}

\begin{abstract}

Conventional antennas, which are widely employed to transmit radio and TV signals, can be used at optical frequencies as long as they are shrunk to nanometer-size dimensions. Optical nanoantennas made of metallic or high-permittivity dielectric nanoparticles allow for enhancing and manipulating light on the scale much smaller than wavelength of light. Based on this ability, optical nanoantennas offer unique opportunities regarding key applications such as optical communications, photovoltaics, non-classical light emission, and sensing. From a multitude of suggested nanoantenna concepts the Yagi-Uda nanoantenna, an optical analogue of the well-established radio-frequency Yagi-Uda antenna, stands out by its efficient unidirectional light emission and enhancement. Following a brief introduction to the emerging field of optical nanoantennas, here we review recent theoretical and experimental activities on optical Yagi-Uda nanoantennas, including their design, fabrication, and applications. We also discuss several extensions of the conventional Yagi-Uda antenna design for broadband and tunable operation, for applications in nanophotonic circuits and photovoltaic devices.
\\
\\
Keywords: optical nanoantennas, Yagi-Uda antennas, all-dielectric nanoantennas, plasmonics, nanoparticles, spectral tuning, Purcell factor

\end{abstract}

\maketitle 

\section{Introduction}

Antennas are all around in our modern wireless society: they are the front-ends in satellites, cell-phones, laptops and other devices that establish the communication by sending and receiving radio waves having frequencies from $300$ GHz to as low as $3$ kHz. However, according to Maxwell's equations, the same principles of directing and receiving electromagnetic waves should be working at various scales independently of the wavelength. Thus, one may naturally ask ''Can a TV-antenna send a beam of light?'' And the answer is ''Yes, optical antennas can!''

The study of optical nanoantennas \cite{bha09,nov11,gia11,bia12} are one of the most promising areas of activity of the current research in nanophotonics due to their ability to bridge the size and impedance mismatch between nanoemitters and free space radiation \cite{sch05}, as well as manipulate light on the scale smaller than wavelength of light \cite{gra10}. Such devices possess collective oscillations of conduction electrons of metals known as plasmon modes, which increase light coupling from nanoemitters to the nanoantenna or from the nanoantenna to freely propagating light, and vice versa. These intriguing properties implicate great potential for the development of novel optical sensors, solar cells, quantum communication systems, and molecular spectroscopy techniques, in particular, for the emission enhancement and directionality control over a broad wavelength range. However, on the other hand, plasmons are also responsible for energy loss in metals at optical frequencies, thereby making some of the traditional low frequency methods of controlling electromagnetic waves futile.

In this paper, we provide an overview of ongoing research into optical Yagi-Uda nanoantennas -- an optical analogue of the well-established radio-frequency (RF) Yagi-Uda antenna, which stands out by its simplicity and excellent directive properties. By discussing and comparing the physics behind electromagnetic wave propagation in RF and optical range we demonstrate the limits of applicability of classical RF architectures to optical nanoantennas and, conversely, show what optical Yagi-Uda nanoantenna can do and RF Yagi-Uda antennas cannot. Based on this comparative analysis, we overview the most recent advances in the physics of optical Yagi-Uda nanoantennas, as well as fabrication and characterization techniques. Finally, focusing on novel broadband and wavelength tunable optical Yagi-Uda nanoantennas, we summarize the potential applications of optical arrayed nanoantennas and discuss open issues.

\section{History}

\begin{figure*}
\includegraphics[width=15cm]{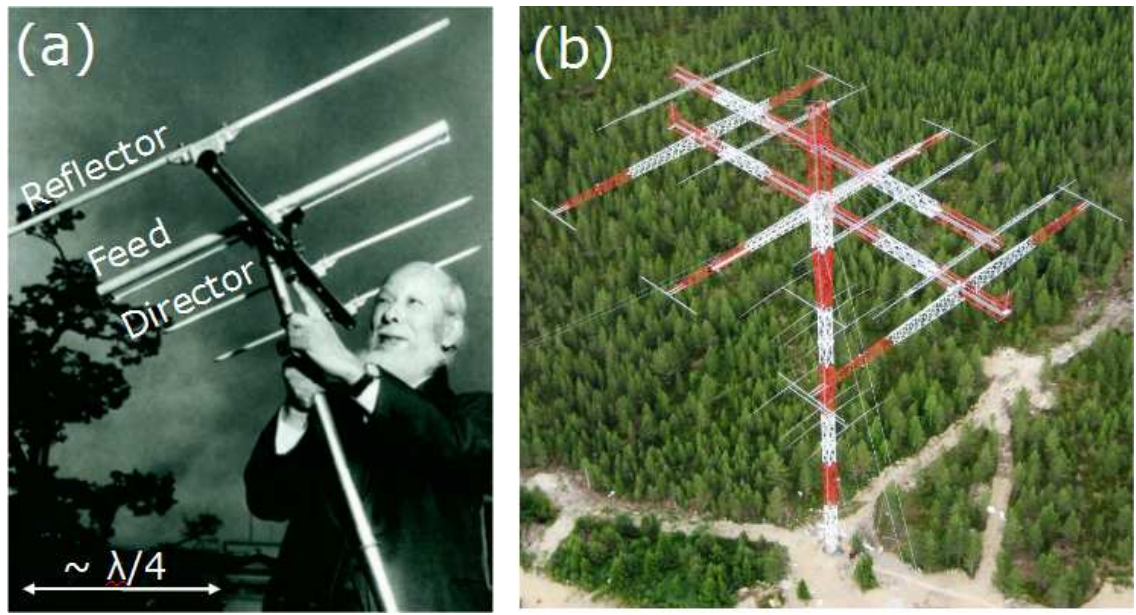}
\caption{(a) Typical RF Yagi-Uda antenna held by one of its inventors Hidetsugu Yagi. The antenna has one reflector and two directors. The arrow shows how the length of the antenna elements is related to the incident wavelength $\lambda$ (reproduced from Ref.~\cite{yagi_nature}). (b) Radio Arcala ''Mammoth'' $160/80$ meter directional Yagi-Uda antenna built in Finland by amateur radio operators. Each of the four guy wires of the antenna extends $120$ meters from the tower representing an area of $170$x$170$ meters. The total tower height and the antenna weight are $100$ m and $39.6$ tons, respectively. Reproduced with permission from www.radioarcala.com.}
\end{figure*}

Recent advances in nanophotonics confirm that the phrase ''There is nothing new except what is forgotten'' \cite{wiki} is also relevant to novel optical antennas that can receive light waves in the same way as car antennas can pick up radio waves. Indeed, it turns out that most of the classical RF antenna design concepts remain applicable at the nanoscale \cite{bha09,nov11}. Moreover, the development of optical nanoantennas faces similar problems as the pioneering experiments with RF antennas in the last decade of the $19$th century, including, the major one -- the weakness of the signals to be received by the antenna.

A simple, efficient, and low-cost solution of this problem for RF antennas was found by a Japanese inventor Shintaro Uda, assistant to Prof. Hidetsugu Yagi of Tohoku Imperial University, Sendai, Japan, in 1926 [Fig. $1$(a)]. He suggested to use multiple single elements to form an array to achieve constructive interference of radio waves in one direction and destructive in the opposite. Thanks to this effect the new antenna was capable of capturing weak RF signals \cite{uda_orig}. Moreover, by virtue of electromagnetic reciprocity, it could also emit RF signals in a directed manner.

The construction turned out to be so useful that, after the translation of Uda's paper from Japanese to English and applications for patents by Yagi (that is why many sources omit the name of Uda), the new antenna design, also known as ''wave channel'', was quickly recognized in the UK, USA, Germany and Soviet Union and implemented both for the transmission and detection of RF waves on airborne radars during the Second World War. Ironically, the invention received just a little attention in Japan \cite{yagi_nature,sat91}.

Nowadays, the conventional Yagi-Uda antenna design, which can be found on almost every roof, not only remains the reference for the ham radio [see Fig. $1$(b)] and TV antennas \cite{bal05}, but also generated a new thrust of activities in nano-optics \cite{cur10,kos10,dre11,dor11} aimed at achieving broadband unidirectional light emission and detection at the nanoscale.

\section{Optical Yagi-Uda nanoantennas}

\begin{figure}
\includegraphics[width=6.5cm]{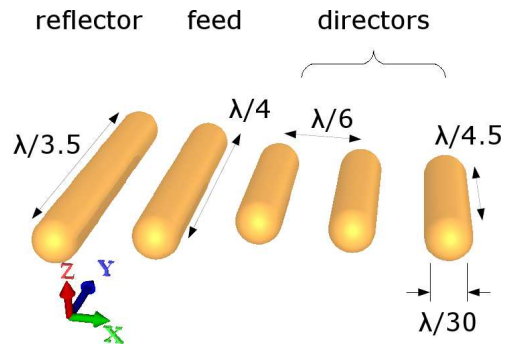}
\caption{Schematic of an optical Yagi-Uda antenna consisting of gold nanorods used for the feed, reflector and directors with the possible dimensions in comparison with the incident wavelength $\lambda$. At the typical telecommunication wavelength of $\lambda=1550$ nm the area occupied by the nanoantenna is $\sim400$x$400$ nm and its weight is $\sim70$ femtogrammes.}
\end{figure}

The Yagi-Uda antenna is one of the most brilliant ideas of directive antennas, that is antennas radiating greater power in only one direction. It is simple to build, and it has a high efficiency in terms of directivity \cite{bal05} and losses that occur throughout the antenna while it is operating at a given frequency. Moreover, it can be shrunk to be as small as $1/25$ of a human hair with a minimum of changes in the construction in order to operate in the infrared and visible regions.

A generic Yagi-Uda antenna consists of an actively driven element called \textit{feed} surrounded by passive elements, which are not driven [Fig. $1$(a)]. These elements are usually of two types: a resonant \textit{reflector} and a few equally spaced \textit{directors}. The key parameter of the whole structure is the length of the feeding element, which should be at resonance at the desired frequency. The lengths of the reflector and directors are chosen to be off resonance and designed to add up in phase in the forward direction, and to cancel in the reverse direction. For that purpose, the director elements are made shorter and the reflector is made longer than the feeding element, which leads to in-- and out-- of-phase current distributions.

\subsection{Plasmonic nanoantennas}

The choice of the optimal length of the feed for optical Yagi-Uda antennas (see Fig. $2$) differs from the standard procedure adopted in RF antenna design \cite{bal05}. Metals at RF are highly conductive, and, thus, can be considered as nearly perfect reflectors. The penetration length of the fields (the skin depth) is negligible compared with the dimensions of the antenna elements. However, at optical frequencies the skin depth is of the order of $10$ nm, and, thus, is comparable with the dimensions of the nanoantenna elements. Therefore, the applicability of classical principles of the RF antenna design principle relying on an incident wavelength is reduced, and it must be revised by rigorously treating a metal as a strongly coupled plasma. The resulting reduced effective wavelength $\lambda_{\rm{eff}}$ seen by optical nanoantennas is related to the incident wavelength $\lambda$ by a simple relation \cite{nov07,bha09,nov11}:

\begin{eqnarray}
\lambda_{\rm{eff}}=a+b\left(\frac{\lambda}{\lambda_{\rm{p}}}\right)
\label{eq:eleven},
\end{eqnarray}

\noindent where $a$ and $b$ are some coefficients with dimensions of length that depend on geometry and material properties \cite{nov07}, and $\lambda_p$ is the plasma wavelength. According to this wavelength scaling rule, an optical half-wave nanoantenna is not $\lambda/2$ in length but is of a shorter length $\lambda_{\rm{eff}}/2$. The difference between the length of the optical and RF dipolar antennas depends on the antenna geometry and is usually in the range of $2-6$ for gold and silver nanoparticles \cite{nov07,bha09}.

The choice of the cross-sectional dimensions of optical nanoantenna elements and internal edge-to-edge spacing between them also differs from the standard RF procedure. In the RF regime, the inter-element spacing typically ranges between $0.1\lambda$ and $0.315\lambda$ and is conditioned by practical rather than physical considerations such as the simplicity and stability of the antenna construction. As a rule, the diameter of RF antenna elements is small compared to the spacing between the elements. At optical wavelengths, however, the antenna element diameter becomes a crucial parameter because the inevitable absorption losses in metals can be reduced by making the volume of the elements larger \cite{hof07}.

Another difference between optical and RF antennas is the excitation of the feeding element. In RF region, Yagi-Uda antennas are locally driven at the feed-gap by electrically connecting the feed, which is typically a $\lambda/2$ dipole, to the feed-line. However, most of the optical nanoantennas investigated so far have been driven from the far-field \cite{nov11}. Such excitation scheme is not optimal for the Yagi-Uda design because the feed is surrounded by other elements placed at small distances as compared with the operating wavelength. Therefore, it would be advantageous to place the light source in the near-field region of the feed. It makes the optical Yagi-Uda nanoantennas perfect candidates for controlling emission of isolated quantum nanoemitters, such as, e.g., quantum dots \cite{shi07} and color centres in nanodiamonds \cite{cas10}.

In order to obtain a strong near-field coupling of an isolated emitter to the fundamental mode of the feed, it is critical to place the emitter at a high electric mode density point. For the nanoantenna geometry shown in Fig. $2$ this point corresponds to one of the ends of the feeding element. A judicious engineering of the feed, such as, e.g., the introduction of a small ($30-50$ nm in width) gap in the centre of the feed, can further increase the coupling strength by exploiting the Purcell effect \cite{purcell_review}, which ensures preferential emission into a highly-confined fundamental cavity mode at high rate.

\begin{figure}
\includegraphics[width=6.5cm]{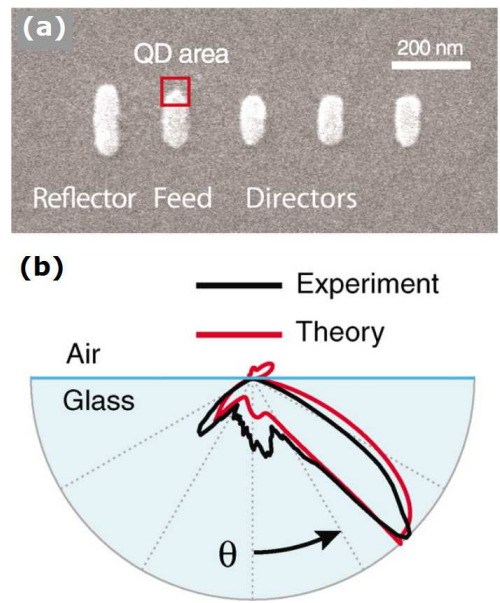}
\caption{Optical Yagi-Uda nanoantenna driven by quantum dots. (a) Scanning electron microscopy image of the fabricated $5$-element Yagi-Uda nanoantenna. (b) Angular radiation diagram for the fabricated Yagi-Uda nanoantenna (black line) vs. theoretical prediction (red line). Figure is reproduced from Ref.~\cite{cur10}.}
\end{figure}

The effective wavelength design principle was used for the fabrication and experimental demonstration of highly unidirectional emission of a quantum dot coupled to an optical Yagi-Uda nanoantenna \cite{cur10} shown in Fig. $3$(a). One of the most important antenna characteristics –- the angular emission diagram -- was calculated from the experimental Fourier-plane image and agrees well with the result of numerical simulations [Fig. $3$(b)]: the nanoantenna emits mainly into the medium which is optically denser. This result is in good agreement with the prediction that emitters at the surface between two interfaces emit mainly into the medium which is optically denser in the case that the emission direction is close to the critical angle \cite{cur10}.

In contrast to earlier experimental study of similar nanoantennas offering directional far-field scattering of a polarized laser beam \cite{kos10}, the device in Fig. $3$(a) ensures full control of the direction of light emission from single quantum emitters. It represents a decisive step toward the development of subwavelength quantum light sources \cite{koe09} with directive properties of RF antennas not attainable with optical cavities \cite{shi07, purcell_review}. The creation of such sources makes it possible to transmit and receive single and entangled photons \cite{shi07} at subwavelength scale, which is an essential feature for quantum cryptography, communication and computing \cite{bou00}.

\begin{figure}
\includegraphics[width=6.5cm]{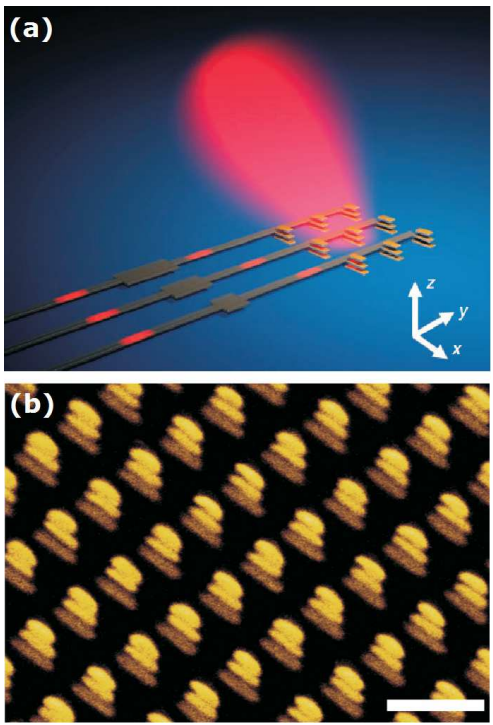}
\caption{Phased array of optical Yagi-Uda nanoantennas. (a) Artist’s view of a 3x3 Yagi-Uda nanoantenna array fed by phase modulating feeding circuits. (b) SEM image of an optical Yagi-Uda nanoantenna array. The gold structure is fabricated on a glass substrate. The nanorods are embedded in photopolymer. Scale bar is 500 nm. Figure is reproduced from Ref.~\cite{dre11}.}
\end{figure}

The near-field excitation design of Yagi-Uda nanoantennas also opens up new ways for the implementation of more complicated RF antenna concepts at the nanoscale. Recently, arrays of three-dimensional ($3$D) optical Yagi-Uda nanoantennas (Fig. $4$), fabricated using top-down fabrication techniques combined with layer-by-layer processing, have been suggested \cite{dre11}. The creation of optical nanoantenna arrays brings the concept of RF phased antenna arrays to the nanoscale, where each antenna is driven independently with a defined phase and amplitude. It considerably widens the applicability range of optical Yagi-Uda nanoantennas, making it possible to steer the direction of the emitted light beam by addressing each nanoantenna independently and controlling the phase of the feeding circuit by means of phase modulators \cite{dre11}.

It is also worth mentioning the efforts targeted to reduce absorption losses in Yagi-Uda nanoantennas by modifying the geometry and the arrangement of the nanoparticles composing the Yagi-Uda architectures \cite{pak09, dev10, bon10, est10, sto11}. These ultracompact nanoantennas look like a Yagi-Uda structure without either reflector or directors. Interestingly, a construction that would lead to a poor performance in RF region not only works well at the nanoscale but also opens up novel opportunities not possible at larger scales, such as a very high front-to-back ratio attainable with only two coupled nanoparticles separated by a distance as small as $\sim1/65$ of the operating wavelength. Another extreme case of a Yagi-Uda structure is represented by nanoantenna arrays consisting of equally spaced metal nanoparticles of the same size \cite{pel09}. These arrays possess Bragg modes allowing for long-range and large area emission enhancement even if the nanoantenna is excited by an emitter with unfavorable dipole moment orientation \cite{pel11}.

\begin{figure}
\includegraphics[width=8.5cm]{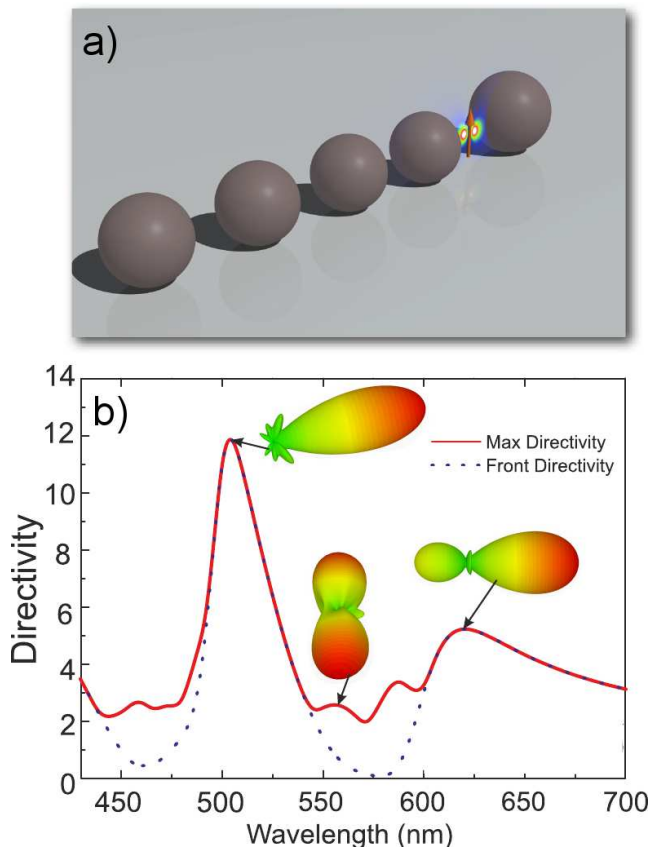}
\caption{ (a) All-dielectric optical Yagi-Uda nanoantenna, consisting of the reflector (sphere 1, right) of the radius of $75$ nm, and smaller directors (spheres 2-5) of the radii of $70$ nm. The dipole source is placed equidistantly between the reflector and the first director at the distance $G$. The separation between the neighboring directors is also $G$. (b) Directivity of the all-dielectric Yagi-Uda nanoantenna as a function of the wavelength for the separation distance $G = 70$ nm. Insets show radiation pattern diagrams at particular wavelengths.}
\end{figure}

\subsection{All-dielectric nanoantennas}

Another way for reducing losses and thereby increasing the antenna efficiency is to utilize high-permittivity low-loss dielectric nanoparticles instead of metallic ones \cite{kra11}. This approach has been already used in RF region \cite{kis05} where Yagi-Uda dielectric resonator antennas are known to be broadband in frequency as compared with all-metal constructions. 

However, in the optical region, the use of dielectric materials promises great advantage over highly absorptive metals. Many studies concentrate on the nanoparticles made of silicon -- one of the key materials of integrated photonics. The real part of the permittivity of the silicon is about $16$ whereas the imaginary part, indicating dissipative losses, is up to two orders of magnitude smaller than that of silver or gold \cite{palik}.

It has recently been suggested to make optical Yagi-Uda nanoantennas of all-dielectric elements \cite{kra12}. All-dielectric Yagi-Uda nanoantennas can be considered as an alternative to their metallic counterparts because of low losses and, as was suggested earlier \cite{paper_krasnok_21}, single spherical nanoparticles made of high-permittivity dielectrics may support both electric and magnetic resonant modes, which is not attainable with single spherical metal nanoparticles. This feature may greatly expand the applicability of optical nanoantennas for, e.g., detection of magnetic dipole transitions of molecules.

Figure $5$(a) shows an example of all-dielectric Yagi-Uda nanoantenna consisting of four directors and one reflector made of silicon. The optimal performance of the nanoantenna at one operating wavelength can be achieved when the director nanoparticles sustain a magnetic resonance and the reflector nanoparticle sustains an electric resonance. Thus, the radii of the directors should be smaller than the radius of the reflector because the positions of the magnetic and electric resonances shift linearly with the radius of the nanoparticle \cite{kra12}. Consequently, the nanoantenna investigated in Ref.~\cite{kra12} consists of nanoparticles with the radii of $70$ nm and $75$ nm used for directors and reflector, respectively.

Figure $5$(b) plots the directivity spectrum of the all-dielectric Yagi-Uda nanoantenna excited by a dipole source placed equidistantly between the reflector and the first director surfaces at the distance $G = 70$ nm. The insets in Fig. $5$(b) show the radiation patterns at particular wavelengths. One observes a strong maximum at $\lambda \approx 500$ nm. The main lobe of the emission pattern at this operating wavelength is extremely narrow with the beam-width about $40^{\rm{o}}$ and negligible backscattering.

However, the radiation efficiencies of both all-dielectric and plasmonic Yagi-Uda nanoantennas are nearly the same for $G = 70$ nm with the averaged value $70\%$. Although dissipation losses of silicon are much smaller than those of silver or gold, the dielectric particles absorb the electromagnetic energy by the whole spherical volume, while the metallic particles do it at the surface only. As a result, there is no substantial difference in the overall performance of these two types of nanoantennas for relatively large distances between the elements. 

The difference between all-dielectric and plasmonic nanoantennas becomes very strong for smaller spacings between the elements: the radiation efficiency of all-dielectric nanoantennas is insensitive to the separation distance whereas the radiation efficiency of plasmonic nanoantennas drops significantly. This is caused by local field enhancement between two adjacent metallic surfaces, which leads to effectively larger absorption of the whole structure. Thus, the use of all-dielectric nanoparticles without the field enhancement allows making more compact nanoantennas e.g. for on-chip wireless communication. But, in contrast to plasmonic nanoantennas, all-dielectric nanoantennas do not provide large spontaneous emission enhancement, which is extremely important for emerging applications such as energy conversion, sensing and quantum information.

\section{Broadband nanoantennas}

\begin{figure*}
\includegraphics[width=15cm]{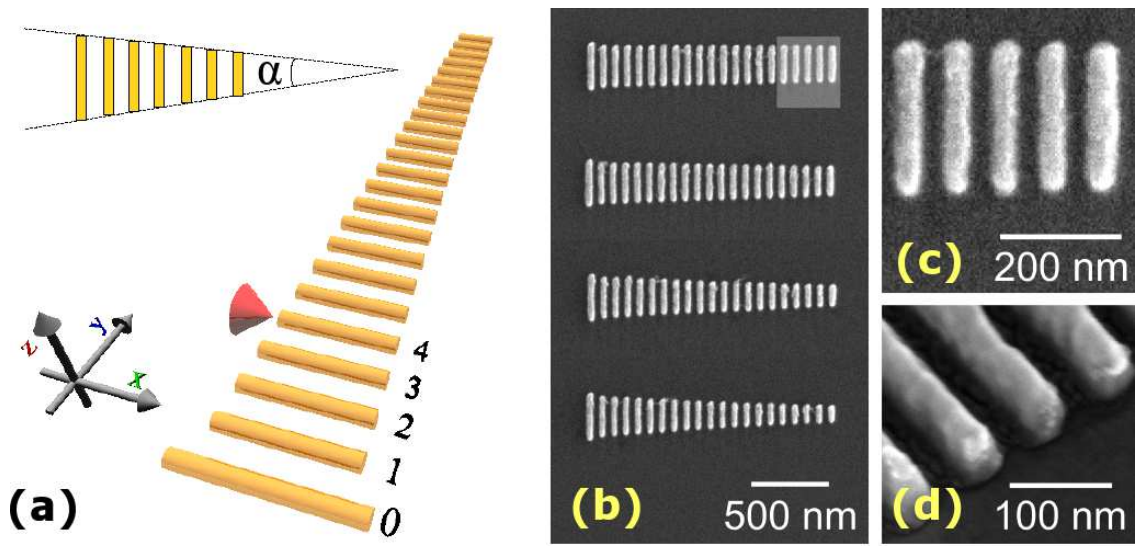}
\caption{(a) Schematic of tapered Yagi-Uda nanoantenna excited by an emitter via one of the possible excitation sites. The inset shows the definition of the tapering angle $\alpha$. (b) Scanning electron microscopy images of the fabricated nanoantennas with different nanorod length variations. (c, d) Close-ups of the highlighted region of one of the nanoantennas \cite{cleo}.}
\end{figure*}

Classical Yagi-Uda antennas are designed to operate at a certain frequency. In RF region, in their basic form Yagi-Uda antennas are very narrowband, with typical bandwidth of $\approx5\%$. Using larger diameter antenna elements, the bandwidth can be extended up to $\approx10\%$.

RF Yagi-Uda antennas can be designed to operate effectively at multiple frequencies. This can be achieved, for example, by creating electrical breaks along each element at which point a parallel LC (inductor and capacitor) circuit is inserted \cite{bal05}. This circuit has the effect of truncating the element at higher frequency band, making it approximately a half wavelength in length. At lower frequencies, the entire element (including the inductance of the LC circuit) is close to half-wave resonance, implementing a different Yagi-Uda antenna. The use of LC circuits presents a very effective and low-cost solution, which is, however, not without disadvantages: the bandwidth and electrical efficiency of the antenna are reduced.

Optical Yagi-Uda nanoantennas demonstrated so far \cite{cur10,kos10,dre11,dor11} are single-frequency and narrowband. However, multifrequency broadband light control is essential for practical applications and has recently become the subject of intensive research \cite{luu12,bor10}. The aforementioned RF strategies for extending bandwidth of Yagi-Uda antennas cannot be applied at optical wavelengths due to two main reasons: (i) the diameter of optical antenna elements is effectively larger that in RF due to fabrication constraints \cite{bha09} and  (ii) feasible optical analogues of inductors and capacitors are not yet available at the nanoscale \cite{alu08}.

Another example of a broadband RF antenna from the Yagi-Uda family is a log-periodic antenna consisting of many dipole elements of decreasing length, all of which can play the role of the feeding element \cite{bal05}. As the dipoles are of different lengths, only one of them is resonant at a given frequency. The bandwidth of log-periodic antennas is relatively wide and is usually two times greater than that of Yagi-Uda antennas. Consequently, the log-periodic design is very attractive for the implementation in the optical and infrared regions.

\subsection{Tapered nanoantennas}

Hereafter, we discuss emerging strategies for the development of optical Yagi-Uda nanoantennas able to receive and emit light waves of different lengths with nearly the same efficiency. As in the preceding sections, we will use and extend ideas known from RF antenna technology.

In order to fully explore the potential of the Yagi-Uda architecture one increases the length of the antenna by incrementing the number of directors. The additional directors help to keep an antenna focused on what is in front of it only. It increases the antenna efficiency when the antenna looks toward the transmitter and ignores signals coming from other directions. 

While the use of this method does not present significant difficulties at RF frequencies, at optical frequencies absorption losses impose severe restrictions on the length of the antenna. Luckily, this restriction can be overcome by slowly tapering the length of the directors. The improvement of the directivity of RF Yagi-Uda antennas by slowly varying the length of directors was suggested by Sengupta \cite{sen60} in $1959$. His idea was abandoned in the modern RF antenna technique because the suggested tapering can be applied to long Yagi-Uda antennas only and results in heavy and unstable constructions. Conversely, at optical frequencies tapering of the plasmonic waveguides \cite{sto04} and metamaterials \cite{roc09} is a very efficient way for nanofocusing of light. The effect of the taper shape and the optimal taper angle on nanofocusing has been discussed actively in the literature (see, e.g., Ref.~\cite{dav10}). 

Recently, it has been demonstrated that tapering of a plasmonic Yagi-Uda nanoantenna [see Fig. $6$(a)] not only serves to focus light and to increase the overall efficiency, but also allows for shrinking down the nanoantenna longitudinal dimension by decreasing the spacing between the elements by a factor of $10$. A pronounced improvement has been found for a particular configuration with $42$ nanorods and the tapering angle $\alpha=6.6^{\rm{o}}$ (see the inset in Fig. $6$(a) for the definition of $\alpha$) at the principal design wavelength $1.515$ $\mu$m \cite{mak11,mak12,mir11}.

Figures $6$(b--d) show experimental tapered Yagi-Uda nanoantennas consisting of $21$ nanorods with different length variations \cite{cleo}. The nanoantennas were fabricated on a glass substrate covered by $5$ nm of indium tin oxide (ITO) using electron-beam lithography followed by evaporation of $50$ nm of gold and a lift-off procedure (see the discussion in the ''Fabrication and Characterization'' section). With lateral nanorod diameters of $\approx 45$ nm and inter-element spacings of $\approx 35$ nm the dimensions of the experimentally realized structures come very close to theoretical design parameters \cite{mak11,mak12,mir11}, rendering them suitable for optical near-field and far-field characterization.

Each nanorod can be used for the excitation of the tapered Yagi-Uda nanoantenna by quantum emitters matched spectrally with the nanorod resonant frequency and placed in the nanorod near-field region \cite{mak12,mir11}. Multifrequency broadband operation of the nanoantenna resulting from an overlap of the spontaneous emission spectra for different emitter positions [Fig. $7$(a)] opens up novel opportunities for broadband emission enhancement, spectroscopy and sensing. A collective excitation with narrow emission linewidth emitters converts the nanoantenna into an efficient multifrequency plasmon-enhanced light ''super-emitter'' with a discrete spectrum [Fig. $7$(b)]. Due to the reciprocity principle, the same tapered nanoantenna can be used both as a transmitter and/or as a receiver, thus being useful for creating a broadband wireless communication system.

\begin{figure}
\includegraphics[width=6.5cm]{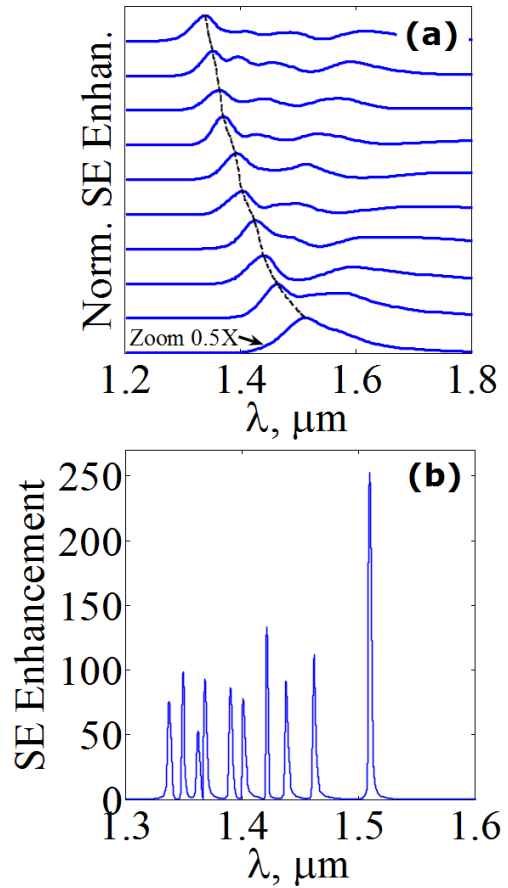}
\caption{(a) Spontaneous emission enhancement spectra of the tapered Yagi-Uda nanoantenna excited with an \textit{x}-polarized emitter via different excitation sites (see Fig. $6$(a) for the definition of the polarization and excitation site numbers). The dashed line, a guide to the eye, connects the maxima of the spectra. (b) Spontaneous emission enhancement spectra of the nanoantenna excited simultaneously by $10$ narrowband x-polarized nanoemitters. Figure is taken from Ref.~\cite{mak12}.}
\end{figure}

\subsection{Log-periodic nanoantennas}

\begin{figure}
\includegraphics[width=6.5cm]{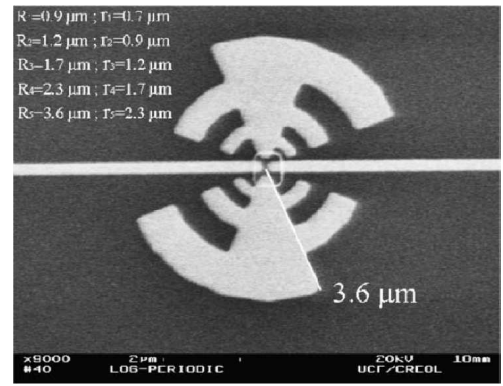}
\caption{Log-periodic optical antenna coupled to a microbolometer. Reproduced from Ref.~\cite{gon05}.}
\end{figure}

Tapered Yagi-Uda nanoantennas are conceptually similar to log-periodic optical nanoantennas inspired by RF log-periodic architectures \cite{bal05}. The original idea of optical log-periodic antennas probably belongs to Gonzalez and Boreman \cite{gon05} who used a lithographic antenna to couple infrared radiation into a bolometer with sub-micron dimensions (Fig. $8$). The excitation of an optical log-periodic antenna with quantum dots has recently been demonstrated experimentally \cite{pavlov_thesis}. In analogy to RF antennas \cite{bal05}, different log-periodic architectures have been considered including arrays of nanorods, nanodipoles and zig-zags.

However, optical log-periodic antennas demonstrated so far just mimic the possible geometries of RF log-periodic antennas such as the length and the spacings of the elements. What has not yet been achieved in the optical range is the excitation of alternating elements with $180^{\rm{o}}$ of phase shift from one another. In RF region, it is normally done by connecting individual elements to alternating wires of a balanced transmission line in order to ensure that the phasing of the elements is correct. Furthermore, as compared with tapered designs, the realization of optical log-periodic nanoantennas requires higher fabrication precision because both the length and the spacings of the elements shrink toward the nanoantenna front-end.

\section{All-optical tunability}

Modern RF antennas are highly optimized in terms of their bandwidth and size, and also can be tuned by means of mechanical reconfiguration of antenna elements or by driving antennas with electronically switchable circuits. In general, spectral tunability allows a smaller antenna to behave as a larger antenna or as an array of antennas \cite{bal05,dre11}, both saving space and improving performance. For example, the arms length of today's ''rabbit ears'' digital TV antennas can be changed in order to tune to the favorite station and deliver a clear high-definition picture. These old-fashioned TV antennas, which appeared in our homes in the 1950's, are very popular because they are compact and can substitute several large roof antennas.   

Consequently, a large and growing body of research investigates tunable optical nanoantennas \cite{far05,hua10,alu08,lar10,abb11,she11,mak11_oe,ala11}. Many novel control mechanisms try to exploit the concept of metamaterial-based \cite{zio08} and non-Foster impedance matching circuits \cite{sus09}, where one of the possible ways for achieving spectral tuning consists in the use of tunable nanocapacitors and/or nanoinductors \cite{eng05,eng_science}. Other approaches rely on vanadium oxide tunable metamaterials \cite{seo10}, mechanically reconfigurable photonic metamaterials \cite{ou11} or metamaterials hybridized with carbon nanotubes \cite{nik10}. Large spectral tunability can also be obtained using electrically controlled liquid crystals \cite{kos05,ber09,ang11}, but a very slow response of liquid crystals is not suitable for many application and, in general, a solid-state implementation is more suitable for on-chip integration of nanoantennas.

However, the spectral tuning of optical Yagi-Uda nanoantennas has so far not been demonstrated because their capability to tune to multiple operating frequencies is compromised by their ability to receive and transmit light in a preferential direction. Tunable Yagi-Uda nanoantennas could have technological applications in building broadband optical wireless communication systems,  advanced nano-sensor systems, high performance solar cells as well as wavelength tunable single photon sources and detectors.

\begin{figure}
\includegraphics[width=8.5cm]{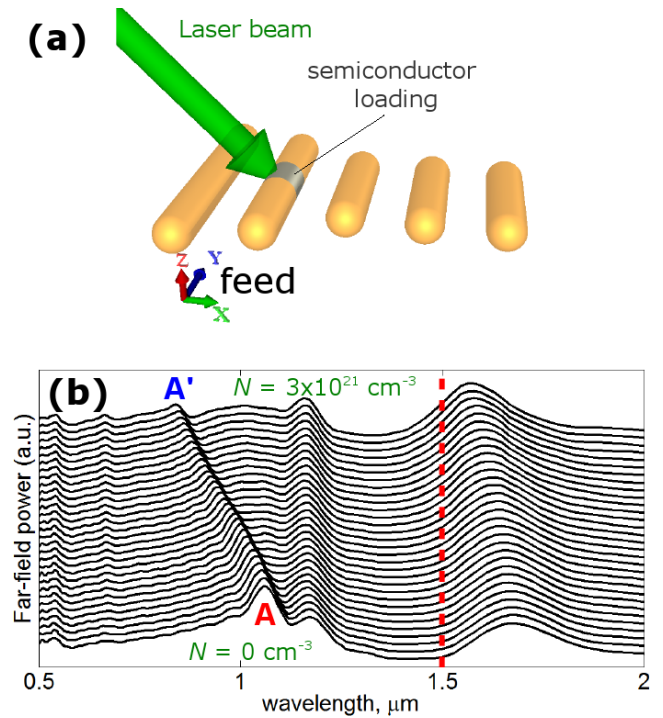}
\caption{\label{fig:epsart} (a) A tunable plasmonic Yagi-Uda nanoantenna consisting of plasmonic nanorods. The gray area of the feed corresponds to the semiconductor nano-disk used as loading. The green arrow schematically shows the direction of the pump laser beam. (b) Far-field power spectra of the nanoantenna in the maximum emission direction as a function of the wavelength for the free carrier densities in the semiconductor loading from $0$ cm$^{-3}$ to $3\cdot10^{21}$ cm$^{-3}$ (from bottom curve and up). Red dashed line indicates the operating wavelength of a Yagi-Uda nanoantenna with the same geometry but equipped with an all-metal feeding element.}
\end{figure}

It has recently been suggested to exploit the free carrier nonlinearity of semiconductors for a dynamical tuning of the operating wavelength of a plasmonic unidirectional Yagi-Uda nanoantenna consisting of plasmonic nanorods used for the feeding element, reflector and directors \cite{mak12_1}. The feeding element of the nanoantenna differs from that in previous designs \cite{cur10} and consists of a semiconductor nano-disk squeezed by two identical nanorods, as shown in Fig. $9$(a). The illumination of the feeding element with a laser beam modifies the free carrier density in the semiconductor and changes the conductivity of the nano-disk. This makes it possible to monotonically tune the operating wavelength of the nanoantenna in a wide spectral range as compared with conventionally designed Yagi-Uda nanoantennas \cite{cur10,kos10,dre11,dor11}. 

Fig. $9$(b) shows the power emitted by the tunable Yagi-Uda nanoantenna in the maximum emission direction as a function of the wavelength of the pump laser beam for different free carrier densities $N$ in the semiconductor loading. We see that the increase in the free carrier density produces a monotonic tuning of the operating wavelengths from $1.07$ $\mu$m [denoted as A in Fig. $9$(b)] to $0.84$ $\mu$m (denoted as A'). 

Moreover, the power emitted by the nanoantenna can form closed bistability loops at different operating wavelengths \cite{mak12_1}. The formation of closed loops was observed earlier in photonic systems exhibiting nonlinear Fano-Feshbach resonances resulting from the interaction between two Fano resonances located very close to each other \cite{mir09}. The closed bistable loops observed may have appealing applications in realizing ultra-fast all-optical switching devices at the nanoscale since the nanoantenna exhibits two different stable states for the same applied optical intensity.

\section{Fabrication and characterization}

In this section, we overview the techniques used to fabricate the Yagi-Uda nanoantennas discussed in the preceding sections. We also survey the experimental techniques used to investigate the optical properties of these nanoantennas in the near-- and far--field zones.

\subsection{Fabrication}

Optical Yagi-Uda nanoantennas are commonly fabricated by standard electron-beam lithography (EBL) followed by metal evaporation and a liftoff procedure \cite{cur10,kos10,dre11}. In this way fabrication accuracies below 50 nm can be obtained. However, this method only allows for the realization of planar antenna geometries. In order to obtain stacked structures like the three-element Yagi-Uda antenna arrays shown in Fig. $4$(b) \cite{dre11}, a multi-step EBL process has to be applied, where several functional layers of metal nanostructures are processed on top of each other. For the success of this method accurate planarization and alignment procedures are crucial. In Ref.~\cite{dre11}, as a first step, gold alignment markers with a thickness of $250$ nm are fabricated on a quartz substrate. Next, a first planar $30$-nm gold nanorod layer is processed using the standard fabrication procedure for in-plane nanostructures. Afterwards, a $100$-nm thick spacer layer consisting of the solidifiable photo polymer PC403 (JCR) is spin-coated on top of this layer, followed by a pre- and a hard-bake. In the next step the alignment markers are used for accurately stacking a second nanorod layer on top of the first one. Finally the planarization, alignment, and in-plane fabrication steps are repeated in order to obtain a third stacked layer.

Similarly, a two step EBL process enables the controlled placement of quantum emitters at defined positions with respect to the nanoantenna as demonstrated in Ref.~\cite{cur10}. In the first lithography step Yagi-Uda nanoantennas are defined on a glass substrate, followed by thermal evaporation of a $30$-nm gold layer and lift-off. Next, a second lithography step is used to define $70$-nm square areas at the hot spots of the antenna feed elements for the formation of a self-assembled monolayer of mercaptoundecanoic acid. Core-shell quantum dots can then be immobilized on these functionalized areas, and in a last step the unexposed resist is removed. The overall result of this procedure is displayed in Fig. $3$ (b).

Other top-down techniques which have been employed for nanoantenna fabrication include focused ion beam (FIB) milling \cite{hua10_1}, nanoimprint lithography \cite{bol09}, and nanostencil lithography \cite{aks10}. While FIB stands out by its versatility to allow for high-precision structuring of almost any conductive substrate, the latter two techniques aim at large-area high-throughput fabrication \cite{bia12}.

Linear plasmonic nanoantenna arrays, which closely resemble the Yagi-Uda geometry, have furthermore been obtained by the controlled assembly of chemically synthesized gold nanocubes using a nanomanipulator \cite{che11, yan10}. Similar approaches rely on nanomanipulation of plasmonic particles with an atomic-force microscope tip \cite{mer08, sch09}. The growing interest in such bottom-up techniques is justified by the purity, and well-defined crystallinity of chemically grown nanoparticles, by the wide range of different nanoparticle shapes and materials/composites available, as well as by the good control of inter-particle distance \cite{bia12}.

\subsection{Characterization}

\subsubsection{Classically designed Yagi-Uda nanoantennas}

\begin{figure*}
\includegraphics[width=12cm]{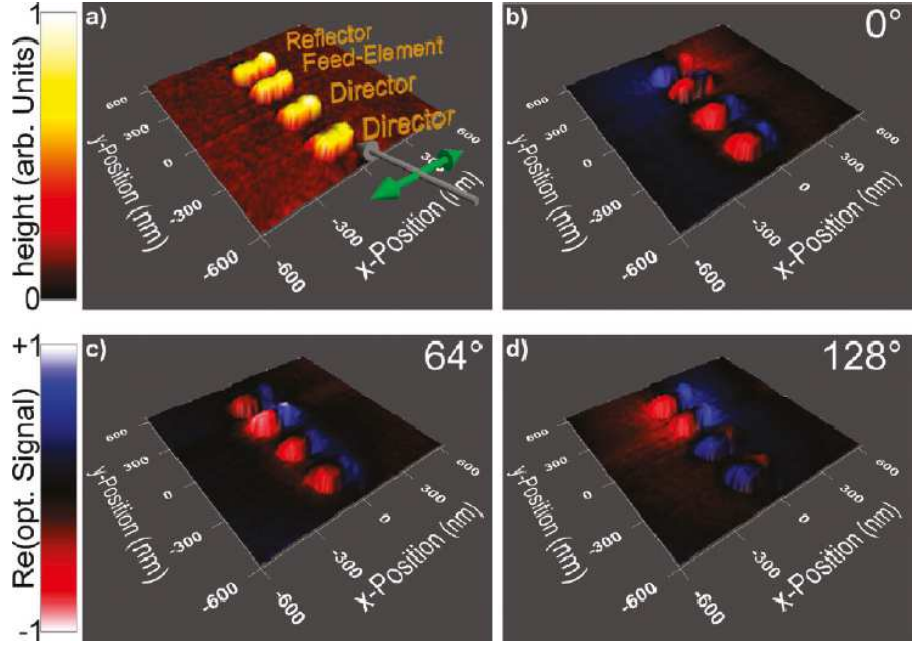}
\caption{(a) $3$D topography of the optical Yagi-Uda nanoantenna. (b-d) Real part of the measured optical signal superimposed onto the topography at different incidents of time: (b) $0^{\rm{o}}$ phase, (c) $64^{\rm{o}}$ phase, and (d) $128^{\rm{o}}$ phase. Figure is reproduced from Ref.~\cite{dor11}.}
\end{figure*}

In analogy to RF antennas, optical nanoantennas can serve both as transmitters and receivers. Because of the reciprocity principle, the emitted radiation pattern of an antenna is equal to its receiving pattern. Experimentally, in RF region it is common to determine the directive properties of antennas by measuring the load current of an antenna receiving radiation from a remote electromagnetic wave source \cite{bal05}. As this technique cannot be used in optical region, several near- and far-field spectroscopic techniques have been developed for the characterization of optical Yagi-Uda nanoantennas \cite{kos10,cur10,dre11,dor11}.

A straightforward way of measuring the optical properties of Yagi-Uda nanoantenna arrays is linear-optical reflectance and transmittance spectroscopy \cite{dre11}. While the antenna array's angular radiation pattern cannot be accessed directly this way, the spectral positions of the plasmonic antenna resonances are identified and evidence for directional behavior is given by differences between reflectance spectra in forward direction, i.e. for light incident from the director side, and backward direction, i.e. for light incident from the reflector side.

A very elegant method of actually measuring a nanoantenna's angular radiation pattern is to characterize the antenna in emission mode by back-focal-plane imaging within a confocal microscope \cite{lie04,cur10}. To this end, emission from the antenna into the substrate is collected with a high-NA objective, and the intensity distribution on the objective's back focal plane or Fourier-plane, which contains the directions of emission towards the substrate, is imaged on an electron-multiplying CCD camera.

In order to characterize the nanoantenna's directive properties in emission mode it is crucial to implement a mechanism ensuring that the antenna is only driven at its feeding element, while the reflector and director elements are only excited in a passive manner through coupling to the feed. In Ref.~\cite{kos10} this is accomplished by tilting the feeding element by $45^{\rm{o}}$, making it possible to couple to it with light polarized perpendicular to the emitted polarization of the Yagi-Uda nanoantenna. An alternative route is to drive the feed via carefully aligned quantum dots. In Ref.~\cite{cur10} quantum dots are placed exclusively in the high-field regions of the feed element. They are excited far off resonance with respect to other antenna elements, thereby again leading to a selective excitation of the feed element. Both of these techniques furthermore offer the advantage to allow for separating the fields emitted by the antenna from the exciting fields by means of additional polarization or color filters brought in front of the detector.

Yagi-Uda nanoantennas in receiving mode have just recently been investigated experimentally using cross-polarization apertureless near-field optical microscopy \cite{dor11}. The local out-of-plane electric field components of the sample are imaged by raster-scanning it underneath the tip. Furthermore, the optical amplitude as well as the optical phase are detected using a homodyne amplification scheme, which allows for visualizing the temporal evolution and the dynamics of the reception process.

Figure $10$(a) shows the $3$D topography of the investigated optical Yagi-Uda nanoantenna fabricated on a glass substrate and illuminated by a weakly focused s-polarized (green arrow) laser beam ($\lambda = 1064$ nm) under an oblique incident angle (grey arrow) \cite{dor11}. Panels (b)-(d) in Fig. $10$ show the real part of the \textit{E}-fields measured above the nanoantenna at different snapshots in time denoted as $0^{\rm{o}}$ phase, $64^{\rm{o}}$ phase and $128^{\rm{o}}$ phase, respectively.

When the phase is of $0^{\rm{o}}$ [Fig. $10$(b)], the two directors show one positive and one negative field lobe with a field node in between them. The field element shows a weak amplitude, and the reflector shows a dipole pointing in the opposite direction with respect to the directors. When the phase is of $64^{\rm{o}}$ [Fig. $10$(c)], all nanoantenna elements show the same colour scheme corresponding to the same dipole moment orientations. Notice that the feed element is at its maximum field, brighter that all the other elements at any time. At $128^{\rm{o}}$ phase [Fig. $10$(d)] the reflector reaches its maximum, but the feed is already declining and the dipole moments of the directors have already flipped.

Thus, the directionality of optical Yagi-Uda nanoantennas manifests itself by the dependence of the local field enhancement at the feeding element on the illumination direction. This effect is similar to the increase in the power measured in experiments with RF antennas receiving signals from a test antenna \cite{bal05}.

Near-field optical measurement of a rotated (backward illuminated) nanoantenna with the same dimensions demonstrated that the change of the illumination direction leads to a destructive interference suppressing the resonance of the feeding element. Therefore, the difference between near-field images for forward and backward illumination of the same antenna structure is also a clear proof of the antenna directionality.

Another experimental technique offering a $\sim10$ nm spatial resolution using a $30$ keV focused electron beam as broad-band point dipole source of visible radiation has been recently presented and applied to study the emission properties of an optical Yagi-Uda antenna composed of a linear array of gold nanoparticles \cite{coe11}. Angle-resolved spectra for different wavelengths shown in Ref.~\cite{coe11} reveal evidence for directional emission of light that depends strongly on where the nanoantenna is excited. These experimental results are explained by a simple and intuitive coupled point dipole model, which includes the effect of the dielectric substrate. In overall, the work establishes angle-resolved cathodoluminescence spectroscopy as a powerful tool to characterize single optical nanoantennas.

\subsubsection{Tapered and log-periodic nanoantennas}

Simulations of tapered \cite{mak11,mir11} and log-periodic \cite{pavlov_thesis} nanoantennas demonstrated that they emit in the forward direction with nearly the same efficiency in almost all possible excitation scenarios. It simplifies their experimental investigations as compared with Yagi-Uda nanoantennas with a single emitter attached to the feeding element only \cite{cur10}. In Ref.~\cite{pavlov_thesis}, a large number of quantum dots was randomly distributed around each log-periodic nanoantenna, leading to a uniform quantum dot distribution with no clustering. These quantum dots excite all nanoantenna elements in an active manner in contrast to the only actively driven feeding element of the classically designed Yagi-Uda \cite{cur10}.

The optical properties of the log-periodic nanoantennas working in emission mode were investigated by back-focal-plane imaging with a confocal microscope \cite{lie04, cur10}. The investigation of the polarization of light emitted by log-periodic nanoantennas with different inter-element spacings revealed that samples with large spacings present the correct polarization transversal to the nanoantenna axis associated with the forward directional emission. However, all nanoantennas were also found to emit light of other polarizations, which differs them from classically designed Yagi-Uda nanoantennas excited by a single nanoemitter and emitting light of only one polarization \cite{cur10}. Far-field radiation patterns of the investigated log-periodic nanoantennas were obtained by processing confocal microscope images and a reasonable agreement between experiment and theory was found.

Interestingly, nanoantennas with small inter-element spacings showed no preferential polarization. This result was explained by a random distribution of quantum dots. Firstly, some quantum dots may contribute to the measured intensity without actually being coupled to the nanoantenna elements. It was confirmed by measuring single nanorods excited by randomly distributed quantum dots. Another reason for not observing directional emission may lie in the enhancement of dipole mode perpendicular to the nanoantenna axis or in higher order multipole mode contribution. Such a mode triggering was demonstrated previously both theoretically and experimentally \cite{chu09}.

\section{Applications}

In this section we discuss the emerging applications of optical Yagi-Uda nanoantennas. We do not pretend to present an exhaustive review, but rather to touch on particularly promising selected topics.

\subsection{Wireless optical communication}

\begin{figure}
\includegraphics[width=8.5cm]{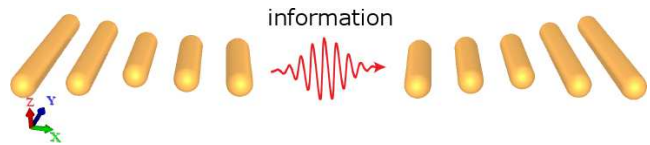}
\caption{Two optical Yagi-Uda nanoantennas can be configured as an on-chip optical interconnect.}
\end{figure}

The objective of optical antenna design is equivalent to that of classical antenna design: to optimize the energy transfer between a localized source or receiver and the free-radiation field \cite{nov11}. In telecommunication, the antenna concept is combined: an antenna can be used both as a receiver and as a transmitter. By utilizing two identical RF antennas, the first one operating as a transmitter and the second one as a receiver, one can create an RF wireless communication system. The same concept can also be used in optics where an on-chip integrated optical wireless system, suggested theoretically in Ref.~\cite{alu10}, might provide an alternative to subwavelength plasmonic waveguides \cite{gra10}, which suffer from absorption losses during light transmission. In analogy to waveguides, nanoantennas confine light below the diffraction limit but, in contrast to waveguides, they transmit it through a channel with relatively low losses.

As the transmitter and the receiver of a wireless system should both have a high directivity, one can envision an optical interconnect consisting of two (or more) optical Yagi-Uda nanoantennas looking one toward another \cite{nov11_1} (Fig. $11$). A quantum dot can be placed in the near field of the nanoantenna-transmitter so that it drives its resonant feeding element. The resulting quantum-dot luminescence will be strongly polarized and highly directed into a narrow forward angular cone, and then received by the nanoantenna-receiver. By placing another nanoemitter near the resonant feed of the nanoantenna-receiver, one can communicate energy to, from, and between nanoemitters. Using a tapered Yagi-Uda \cite{mak11, mir11} or a log-periodic \cite{pavlov_thesis} nanoantenna one can transmit energy to, from and between many nanoemitters at different frequencies simultaneously by precisely placing them near individual nanoantenna elements.

Furthermore, tunable Yagi-Uda nanoantennas with a semiconductor loading of the feeding element \cite{mak12_1} can perform as active components of optical wireless communication systems. They can perform not only as frequency selective transmitters or receivers, but also as an all-optical bistable switching device \cite{gib85} exhibiting emission properties not attainable with actively tunable dipole or bow-tie plasmonic nanoantennas \cite{che10,zho10,seo10}.

\subsection{Nanoscale spectroscopy}

Nanoscale field confinement by optical nanoantennas offers remarkable opportunities for single molecule detection \cite{tam08,ang08,ang10} and nanospectroscopy \cite{sch11}. Yagi-Uda nanoantennas can perform as an optical spectrum analyzer at the nanometer scale that is able to distribute different frequency contents of the radiation of an optical dipole source into different directions in space \cite{li09}. The feasibility of this concept has been demonstrated theoretically based on optical Yagi-Uda nanoantennas composed of plasmonic core-shell nanoparticles. Core-shell plasmonic nanoparticles with the core made of an ordinary dielectric and the shell of a plasmonic material, or vice versa, are ideal for this purpose because their resonant frequencies can be adjusted in a broad range by varying the ratio of the core radius $b$ to the outer radius $a$. Specifically, $b/a$ of each particle can be designed deliberately to ''detune'' its resonance so that they may play the role of the ''reflectors'' or the ''directors'' in a conventional RF Yagi-Uda antenna.

The beam pattern of a core-shell Yagi-Uda nanoantenna is sensitive to the variation of the operating wavelength for two reasons.  Firstly, the relative permittivities of the plasmonic materials forming the nanoparticles, and the resulting electric polarizability of these particles, are wavelength dependent. Secondly, when the sizes of the nanoparticles and their relative positions are decided and then kept fixed, different operating wavelengths would lead to different relative electrical sizes and relative distances, which causes variation in coupling among particles. Therefore, optical nanoantennas composed of plasmonic particles are intrinsically sensitive to wavelength variation. Using this unique property, one can envision an innovative nanoscale ''Yagi-Uda'' spectrum analyzer in which the absorption and emission of a fluorescent molecule can be separated away both spectrally and spatially. 

\subsection{Photovoltaic devices}

\begin{figure}
\includegraphics[width=8.5cm]{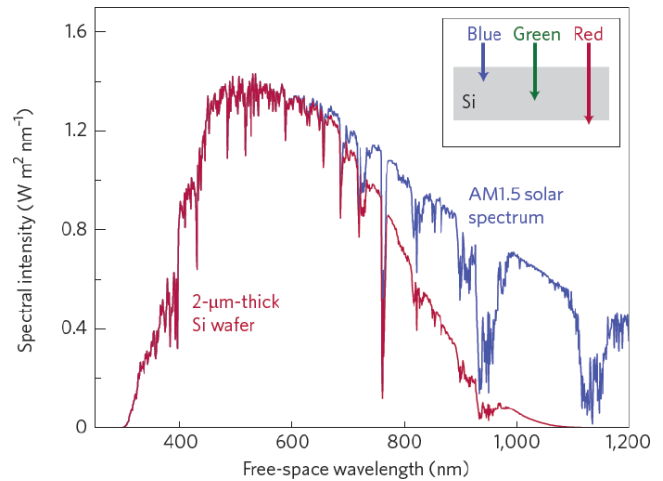}
\caption{AM1.5 solar spectrum, together with a graph that indicates the solar energy absorbed in a $2$-$\mu$m-thick crystalline Si film. Figure is reproduced from Ref.~\cite{atw10}.}
\end{figure}

Photovoltaics, a promising technology relying on the conversion of sunlight to electricity, may allow for the generation of electrical power on a very large scale. It could thus make a considerable contribution to solving the energy problem faced by our society. Plasmonic excitation and light localization can be used advantageously in high-efficient photovoltaics \cite{atw10}.

Conventionally, photovoltaic absorbers are optically thick in order to allow better light absorption and photocarrier current collection. However, a large portion of the solar spectrum, especially in the $600-1100$ nm spectral range, is poorly absorbed (Fig. $12$). The use of plasmonic nanoantennas of subwavelength dimensions can reduce the physical thickness of photovoltaic absorber layers while keeping their optical thickness constant. These nanoantenna-receivers can trap and couple freely propagating sunlight into an absorbing semiconductor thin film, increasing the effective absorption cross-section.

Broadband tapered Yagi-Uda nanoantennas extend the list of advantages of plasmonic nanoantennas for photovolatics because they can be designed to absorb sunlight in a very broad spectral range. In order to achieve this specific goal, the number of nanorods and the nanorod length variation can be chosen \cite{mir11, mak12} such that the antenna spectrum covers the AM$1.5$ solar spectrum including the ''problematic'' spectral range of $600-1100$ nm. Since the reception characteristics of the nanoantenna are nearly constant at all operating wavelengths, it is expected that all fractions of the solar spectrum would be absorbed almost equally.

\subsection{Nanosensing}

\begin{figure}
\includegraphics[width=8.5cm]{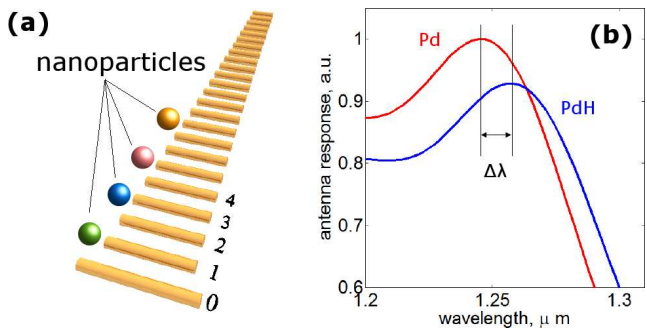}
\caption{(a) Schematic of a broadband plasmonic array Yagi-Uda nanoantenna consisting of metal nanorods of varying length. Active nanoparticles are placed in close vicinity of the nanorods. (b) Optical sensing principle using the nanoantenna interacting with only one active palladium nanoparticle placed at the excitation site $n = 1$. A change in the optical properties of the nanoparticle on exposure to hydrogen (Pd to PdH) causes a resonance shift ($\Delta \lambda$) in the optical response of the nanoantenna.}
\end{figure}

The scientific advances of recent years have recognized optical nanoantennas as a technology that could make possible a range of innovations. One of the important areas where the nanoantennas have immediate implications are plasmon mediated chemistry \cite{che83} and sensing \cite{zha10,aim09,ada09,hao09,kha09,liu11}.

Tapered Yagi-Uda nanoantennas can develop significantly the idea of the nanoantenna-enhanced gas sensing \cite{liu11}. In order to achieve this goal, single nanoparticles made of gas-sensitive active materials such as platinum (Pt), palladium (Pd) or iridium (Ir) can be precisely placed in the near-field region of individual nanorods of nanoantennas as shown in Fig. $13$ \cite{antenna_sensor}. Alternatively, Pt, Pd or Ir can be deposited on the top of the nanoantenna elements. Yet another configuration is a tapered array of elements made of different metals. 

A change in the optical properties of the active nanoparticles will influence the resonances of the individual nanorods, which in turn inevitably changes the far-field characteristics of the nanoantenna at different operating wavelengths. Thanks to a high sensitivity of the Yagi-Uda architecture, the expected spectral shift will be larger as compared with a several nanometers shift observed with single bow-tie nanoantennas \cite{liu11,titl12}. Furthermore, as shown in Fig. $7$(b), the separation between the peaks in the nanoantenna spectra is large enough to allow for easy reading of data from different nanoparticles.

Any change in the far-field response of the nanoantenna can be detected using conventional far-field spectroscopy techniques \cite{nov11} or with another broadband Yagi-Uda nanoantenna employed as a receiver. The latter configuration is attractive for application to horizontal on-chip plasmonic wireless signal-transmission circuits, which can be used for real-time transmission of data monitored by the nanosensor. Since the data on different monitored chemical substances can be transmitted simultaneously at different wavelengths, the suggested nanoantenna-based sensors can be used in the development of nanoscale optical nose devices that can be tuned both geometrically and dynamically to monitor different gases at different operating wavelengths.

\subsection{Quantum communication and information}

Efficient sources of single and entangled photons are crucial for the development of quantum information technology \cite{bou00}. Single semiconductor quantum dots can emit single photons and degenerate polarization entangled photon pairs consisting of photons with the horizontal and vertical polarizations \cite{shi07}. In order to collect the emitted photons, the quantum dot spontaneous emission can be controlled and funnelled in an optical cavity mode. When a quantum dot is spatially and spectrally coupled to a cavity mode, its emission rate into the mode is increased by the Purcell factor $F_{\rm{P}}$ \cite{purcell_review} and a fraction $\frac{F_{\rm{P}}}{F_{\rm{P}}+1}$ of the quantum dot emission is funnelled into the mode, which opens up a possibility of a high collection efficiency of the emitted photons. 

Yagi-Uda nanoantennas can be used for the emission of single photons with a high efficiency as compared with optical cavities because they provide a large and broadband Purcell factor accompanied by a high directivity \cite{cur10}. Moreover, using broadband tapered \cite{mak11,mir11} and log-periodic \cite{pavlov_thesis} nanoantennas one can generate single photons at different frequencies at the same time. However, the implementation of Yagi-Uda nanoantennas to the extraction of polarization entangled photon pairs is challenging because of two different spontaneous emission enhancement rates for the different photon polarizations that eventually erases the entanglement. One of the possible solutions to this problem can be found by fabricating $3$D Yagi--Uda nanoantennas one in close proximity to another. This arrangement also known as photonic molecule \cite{dou10} can offer identical emission conditions for different photon polarizations. 

\subsection{Optical metamaterials}

\begin{figure*}
\includegraphics[width=12cm]{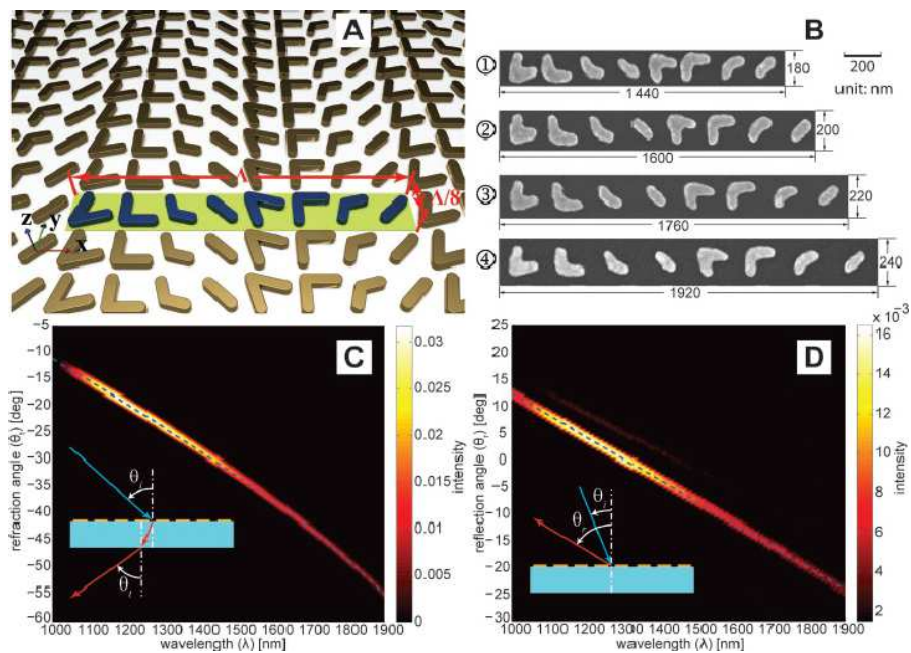}
\caption{(a) Schematic view of the metamaterial nanoantenna array. The unit cell of the plasmonic interface (in blue) consists of eight gold V-shaped nanoparticles of $40$-nm arm width and $30$-nm thickness, repeated with a periodicity of $\Lambda$ in the \textit{x} direction and $\lambda/8$ in the \textit{y} direction. The phase delay for cross--polarized light increases along the \textit{x} axis from right to left. (b) Scanning electron microscope images of the unit cells of four antenna arrays with different periodicities fabricated on a single silicon wafer. (c, d) The false-color maps indicate the experimentally measured relative intensity profiles for the antenna array labeled ($1$) (see panel (b), $\Lambda = 1440$ nm) with \textit{x}-polarized excitation. The dashed line shows the theoretical prediction of the peak position using the generalized Snell's law. (c) Refraction angle versus wavelength for cross-polarized light with $30^{\rm{o}}$ incidence angle. (d) Reflection angle versus wavelength for cross-polarized light with $65^{\rm{o}}$ incidence angle. Figure is reproduced from Ref.~\cite{ni11}.}
\end{figure*}

In the previous sections we have demonstrated that arrays of plasmonic particles forming Yagi-Uda nanoantennas are able to precisely manipulate light in the same way as arrays of metal bars manipulate radio waves. Nowadays, a large and growing body of research investigates novel artificial materials -- metamaterials \cite{veselago} able to manipulate light in new ways not existing in nature. Metamaterials could make possible a range of optical innovations such as, e.g., more powerful microscopes, telecommunications and computers.

One of the approaches for creating metamaterials is to abruptly change the phase of light, which dramatically modifies how light propagates. However, this concept is not new. Indeed, the idea of controlling the phase was used by Uda and Yagi in their famous antenna $85$ years ago. We remind that the reflector of a Yagi-Uda antenna is longer than its resonant length, its impedance is inductive and thus the current on the reflector lags the voltage induced on the reflector. The director elements are shorter than the resonant length, which makes them capacitive, so that the current leads the voltage. This causes a phase distribution to occur across the elements, simulating the phase progression of a plane wave across the array of elements.

However, Uda and Yagi could not know that a long-held Snell's law, used to describe how light reflects and refracts while passing from one material into another, would be modified $85$ years after their invention. It has been recently shown in Ref.~\cite{yu11}, that the phase of light and the propagation direction can be changed dramatically by using metamaterials based on an array of V-shaped plasmonic nanoparticles. What was pointed out in Ref.~\cite{yu11} and demonstrated at one wavelength is revolutionary. Very recently, this idea has been extended to the near infrared, which is essential for telecommunications, and more importantly applied to a broad spectral range from $1$ to $1.9$ $\mu$m \cite{ni11}. The resulting nanoantenna array [see Fig. $14$(a,b)] offers an unparalleled control of anomalous reflection and refraction, including negative refraction [Fig. $14$(c,d)]. This could lead to a variety of applications such as spatial phase modulation, beam shaping, beam steering, and plasmonic lenses, and it could also have an impact on transformation optics and on-chip optics.

\section{Conclusion and perspectives}
We have reviewed the recent advances in the study of unidirectional optical nanoantennas, in particular, the optical analogue of the famous Yagi-Uda antenna architecture. Owing to a great potential of this type of nanoantennas, evidenced by a rapid growth of research activity in this direction the presented review is not meant to be exhaustive, but it gives rather a bird's-eye view of optical Yagi-Uda nanoantennas and some of their applications.

Suggested over $85$ years ago, Yagi-Uda antennas have played a crucial role in the establishment of radar technology, radio communication, television broadcasting, and ham radio. Nowadays, optical analogues of the Yagi-Uda architecture open up unique perspectives and great potentials to explore new avenues in the fields of active nanophotonic circuits, quantum information technology, high-density data storage, optical and bio-sensing, medical imaging systems, photovoltaics and photodetection.
\\
\section*{Acknowledgments}
This work was supported by the Australian Research Council. The fabrication facilities were supported by the Australian National Fabrication Facility (ACT node). The authors thank A. Davoyan, C. Simovski, A. Krasnok, P. Belov, M. Decker, D. Neshev, H.H. Tan and C. Jagadish for valuable contributions to the work on tapered and all-dielectric nanoantennas. The authors acknowledge many useful discussions with their colleagues from the Nonlinear Physics Centre and Metamaterial Meeting Group at the Australian National University.

\end{document}